\documentstyle[twocolumn]{mn}
\newcommand{\nn}{\nonumber \\} 
\newcommand{\be}{\begin{equation}}
\newcommand{\ee}{\end{equation}}
\newcommand{\ba}{\begin{eqnarray}}
\newcommand{\ea}{\end{eqnarray}}
\newcommand{\mincir}{\raise -2.truept\hbox{\rlap{\hbox{$\sim$}}\raise5.truept 
\hbox{$<$}\ }} 
\newcommand{\magcir}{\raise -2.truept\hbox{\rlap{\hbox{$\sim$}}\raise5.truept 
\hbox{$>$}\ }} 
\newcommand{\minmag}{\raise-2.truept\hbox{\rlap{\hbox{$<$}}\raise 6.truept\hbox 
{$>$}\ }}
\input{psfig.sty}
\title[Constraining the cosmological baryon density with X-ray clusters]
{Constraining the cosmological baryon density with X-ray clusters}
\author[Claudio Gheller, Ornella Pantano
and Lauro Moscardini]{Claudio Gheller$^{1}$, 
Ornella Pantano$^{2}$
and Lauro Moscardini$^{3}$ \\
$^{1}$ SISSA -- International School for Advanced Studies,
via Beirut 2--4, I--34013 Trieste, Italy\\
$^{2}$ Dipartimento di Fisica {\em Galileo Galilei}, Universit\`{a} di
Padova, via Marzolo 8, I--35131 Padova, Italy\\
$^{3}$ Dipartimento di Astronomia, Universit\`{a} di Padova, vicolo
dell'Osservatorio 5, I--35122 Padova, Italy\\}

\begin{document}

\maketitle

\begin{abstract}
We study the properties of X-ray galaxy clusters in four cold dark matter
models with different baryon fraction $\Omega_{BM}$ ranging from 5 to 20 per
cent. By using an original three-dimensional hydrodynamic code based on the
piecewise parabolic method, we run simulations on a box with size $64\ h^{-1}$
Mpc and we identify the clusters by selecting the peaks in the X-ray luminosity
field. We analyse these mock catalogues by computing the mass function, the
luminosity function, the temperature distribution and the
luminosity-temperature relation. By comparing the predictions of the different
models to a series of recent observational results, we find that only the
models with low baryonic content agree with the data, while models with larger
baryon fraction are well outside the 1-$\sigma$ errorbars. In particular, the
analysis of the luminosity functions, both bolometric and in the energy band
[0.5--2] keV, requires $\Omega_{BM}\mincir 0.05$ when we fix the values $h=0.5$
and $n=0.8$ for the Hubble parameter and the primordial spectral index,
respectively. Moreover we find that, independently of the cosmological
scenario, all the considered quantities have a very little redshift evolution,
particularly between $z=0.5$ and $z=0$. 
\end{abstract}

\begin{keywords}
Hydrodynamics -- Large-scale structure of the Universe -- X-ray: galaxies,
general -- Dark matter -- Galaxies: clusters --Cosmology: theory 
\end{keywords}

\section{Introduction}

Galaxy clusters are the most extended gravitationally bound systems of the
Universe. In spite of their size, they are simple objects: their dynamics is
governed essentially by the balance between gravitational and pressure forces,
controlled by shock heating and adiabatic compression. Other processes, like
radiative cooling in cluster cores or population III stars heating, may play a
significant role only in some aspects of their history. For these reasons, the
evolution and present day properties of clusters are very sensitive to the
fundamental cosmological parameters and to the initial power spectrum.
Therefore they provide an ideal tool to study the formation of the structures
of the universe. Analytic techniques, as the Press-Schechter (1974) formalism,
and numerical simulations, based on the Zel'dovich (1970) approximation or on
more accurate N-body codes, have been extensively used to put strong
constraints on the cosmological scenarios by comparing the model predictions to
the observed abundances and clustering properties (see e.g. Borgani et al. 1997
and references therein). 

It is well known that clusters have a large X-ray emission, essentially
produced by thermal bremsstrahlung. The quality of the observations in this
band have been recently largely improved, providing new data which potentially
can discriminate between different cosmological models. In order to  compare
these data to the numerical simulations we must include a reliable treatment of
the hydrodynamic quantities, such as temperature, pressure and energy. For
these reasons in the last years there has been a large effort of the
cosmological community to develop new numerical codes which combine the
standard N-body techniques, which solve the dynamics of the dark matter
component, to hydrodynamic methods for properly following the evolution of the
baryonic component (Hernquist \& Katz 1989; Evrard 1990; Cen 1992; Steinmetz \&
Muller 1993; Ryu et al. 1993; Bryan et al. 1995; Gnedin 1995; Pen 1997; for a
comparison of the different algorithms see Kang et al. 1994b). These codes have
been applied for simulating in various cosmological models galaxy cluster
properties and evolution and for calculating their masses, X-ray emission, mean
temperature and the relations between these quantities, that is the fundamental
information that is usually obtained from observational data (see e.g. Kang et
al. 1994a; Bryan et al. 1994a,b; Navarro, Frenk \& White 1995; Pen 1996; Cen
1997). 

In this paper we apply a hydrodynamic code based on the coupling of the
Piecewise Parabolic Method with the Particle-Mesh N-body code (Gheller, Pantano
\& Moscardini 1997) to study the large-scale distribution of the X-ray clusters
in the framework of critical-density cold dark matter (CDM) models with a high
baryon content (up to 20 per cent). This kind of model, allowed by the present
uncertainties in the determination of the baryon density $\Omega_{BM}$ of the
universe, when considered with a small tilt of the primordial spectral index
($n \approx 0.8$), has been suggested as possible solution of the problems of
the standard (i.e. with $\Omega_{BM}\approx 0.05$ and $n=1$) CDM model, namely
the high small-scale power and the low first peak in the microwave anisotropy
power spectrum (White et al. 1996). For a discussion about tilted models,
see also Lucchin \& Matarrese (1985); Vittorio, Matarrese \& Lucchin (1988);
Liddle \& Lyth (1993).

The plan of the paper is as follows. In Section 2 we introduce the cosmological
models that we consider. In Section 3 we present the numerical code and the
method of cluster identification. The results are shown in Section 4. The main
conclusions are drawn in Section 5. 

\section{Cold Dark Matter models with high baryon content} 

For a long period the CDM scenario has been the reference model for the
interpretation of the observational data on the large scale of the universe.
Its standard version assumes a flat universe with a density parameter
$\Omega_0=1$, a Hubble constant $h=0.5$ (in units of 100 km s$^{-1}$
Mpc$^{-1}$), a baryon contribution to the density $\Omega_{BM}$ fixed by the
standard theory of big bang nucleosynthesis (BBN), and primordial fluctuations
with Gaussian distribution and power spectrum $P(k) \propto k^n$, with $n=1$.
However, the normalization implied by the COBE detection of the microwave
anisotropies (Smoot et al. 1992; see also Bennett et al. 1996) changed the
situation, giving a spectrum with too much power on scales smaller than $10\
h^{-1}$ Mpc. As a consequence, the CDM model is not able to reproduce either
the clustering properties of galaxies and the distribution and abundances of
clusters. 

In order to solve the CDM problems, many alternatives have been proposed and
discussed to reduce the short-scale power: tilted CDM models, i.e with a
primordial spectral index $n<1$; mixed dark matter models, i.e. with the
addition of a hot component corresponding to approximately 20 per cent of the
total density; open CDM models, i.e. with $\Omega_0 < 1$; Lambda CDM models,
i.e. with a non-vanishing cosmological constant fixed such as its contribution
to the density is $\Omega_\Lambda=1-\Omega_0$; CDM models with a small Hubble
constant, i.e. with $h<0.5$. For a general review on the CDM model and its
variants, see Liddle \& Lyth (1993) and Coles (1996). 

Very recently, it has been suggested (White et al. 1996) that also an increase
of the baryon content $\Omega_{BM}$ can produce critical-density CDM models
which fit the observational data in a reasonable way. In the past, the mean
baryon density has been considered as a fixed quantity for most of cosmological
models, in agreement with the predictions of the standard theory of BBN and the
local estimates of light elements abundances: $\Omega_{BM} h^2 = 0.0125\pm
0.0025$, where the uncertainty is at 95 per cent confidence level. 
 
The possibility of varying $\Omega_{BM}$ has been introduced because some
recent observations have suggested that its estimate is much more uncertain
than previously thought. In fact the first measurements of the deuterium
abundance at high redshift yield very discrepant values. Small $\Omega_{BM}$,
in good agreement with the BBN predictions, are implied from different
measurements of the neutral hydrogen column density in high-redshift clouds
(Rugers \& Hogan 1996a,b; Songaila, Wampler \& Cowie 1997). On the contrary,
the analyses performed by Tytler, Fan \& Burles (1996) suggest $\Omega_{BM} h^2
= 0.024 \pm 0.002\pm 0.002\pm 0.001$, with the $1\sigma$ uncertainties being
statistical, systematic and theoretical, respectively. These last results are
also consistent with the lower limits obtained by Rauch et al. (1997) by
comparing the observed flux decrement distribution function from a sample of
seven high resolution QSO spectra to simulations of the Ly$\alpha$ forest.
Similar conclusions have been reached also by Weinberg et al. (1997) by
computing the amount of neutral hydrogen present in the high-redshift
intergalactic medium necessary to produce the Ly$\alpha$ absorption in the QSO
spectra. Finally, the determination of the baryonic fraction in clusters of
galaxies (White et al. 1993; White \& Fabian 1995; Elbaz, Arnaud \& B\"ohringer
1995) leads, in a critical-density universe, to high values of $\Omega_{BM}$.
For example White \& Fabian (1995) have calculated: 
\be
{\Omega_{BM}\over\Omega_0}=0.14^{+0.08}_{-0.04}\left({h\over 0.5}\right)^{-3/2}
\ ,
\ee
again at the 95 per cent confidence. 

By using a semi-analytical approach, White et al. (1996) carried out a general
exploration of CDM models with high $\Omega_{BM}$, allowing also the Hubble
parameter $h$ and the spectral index $n$ to vary. The predictions of these
scenarios have been compared with the observations of the clustering properties
of galaxies, the cluster abundances, the statistics of the peculiar velocities,
the formation of high-redshift objects (i.e. damped  Ly$\alpha$ systems, Lyman
break galaxies, quasars and clusters) and the cosmic microwave background
anisotropies. Their final suggestion is that the models with $\Omega_{BM}$ in
the range [0.1--0.2], a Hubble parameter $h\approx 0.5$ and a small tilt in the
primordial spectrum ($n\approx 0.8$), are in good agreement with all these
data. A high baryonic fraction in fact helps ``naturally'' to suppress the
short-scale power, as baryon collapse is stopped till decoupling, and, at the
same time, amplifies the first peak of the CMB spectrum, compensating for the
loss of height introduced by the tilt (needed to avoid too low values of $h$,
excluded by the observational data). At the end of their analysis they conclude
that these models can represent at the moment a viable alternative in the
framework of critical-density CDM models. 

In this paper we consider models where the universe is spatially flat, i.e.
$\Omega_0\ =\ \Omega_{DM}+\Omega_{BM}$ = 1. Here $\Omega_{DM}$ is the density
parameter of the (cold) dark matter component. We assume a vanishing
cosmological constant ($\Omega_\Lambda=0$). The initial spectrum of
perturbations is defined as $P(k)\ \propto\ k^n T^2(k)$, where $T(k)$ is the
CDM transfer function. We use the expression obtained by Bardeen et al. (1986):
\ba
T(q) & = & {\ln \left( {1+2.34q}\right) \over 2.34q}\ \times 
[1+3.89q+ \nn
& & (16.1q)^2+(5.46q)^3+(6.71q)^4]^{-1/4}\ ,
\ea
where $q=k/h\Gamma$. The shape parameter $\Gamma$ takes into account the
dependence on the Hubble parameter $h$, on the total density $\Omega_0$ and on
the baryon density $\Omega_{BM}$ (Sugiyama 1995): 
\be
\Gamma\ =\ \Omega_0 h\exp (-\Omega_{BM}-\sqrt{h/0.5}\ \Omega_{BM}/\Omega_0)\ .
\ee

As suggested by White et al. (1996), we have fixed $h=0.5$ and $n=0.8$,
allowing the baryon content to vary. We consider four different values for
$\Omega_{BM}$: the usual value determined by the standard theory of BBN
($\Omega_{BM}=0.05$); a baryon abundance consistent with the low deuterium
measurements and with the more recent BBN calculations ($\Omega_{BM}=0.1$); a
value close to the estimate resulting from the cluster baryon fraction in the
case of critical-density universe ($\Omega_{BM}=0.15$); a more extreme case
representing the upper limit of the range of observed cluster baryon fraction
($\Omega_{BM}=0.2$). In the following we will label these four models as BM05,
BM10, BM15 and BM20, respectively. The normalization of the spectrum, usually
parametrized by $\sigma_8$, the matter rms fluctuation in a top-hat sphere of
radius $8\ h^{-1}$ Mpc, is defined by the four-year COBE data (Bunn \& White
1997). The cosmological parameters used for the different models are summarized
in Table 1. Since the power at small scales (i.e. on large $k$) increases when
$\Omega_{BM}$ decreases, we can expect a faster evolution in the models with
low baryon content, with a larger production of big mass overdensities. 

\begin{table}
\centering
\caption[]{The parameters of the cosmological models. Column 2: the density 
parameter $\Omega_0$; Column 3: the baryon density $\Omega_{BM}$; Column 4:
the primordial spectral index $n$; Column 5: the Hubble parameter $h$; Column
6: the shape parameter $\Gamma$; Column 7: the spectrum normalization 
$\sigma_8$.}
\tabcolsep 4pt
\begin{tabular}{lcccccc} \\ \\ \hline \hline
Model & $\Omega_0$ & $\Omega_{BM}$ & $n$ & $h$ & $\Gamma$ & $\sigma_8$ \\ \hline
BM05 & 1.0 & 0.05& 0.8& 0.5& 0.44& 0.77 \\
BM10 & 1.0 & 0.10& 0.8& 0.5& 0.41& 0.72 \\
BM15 & 1.0 & 0.15& 0.8& 0.5& 0.39& 0.66 \\
BM20 & 1.0 & 0.20& 0.8& 0.5& 0.36& 0.61 \\
\hline
\end{tabular}
\end{table}
 
\section{Numerical Method}

\subsection{The code}

The simulations have been evolved by using the numerical code presented in
Gheller et al. (1997; see also Gheller, Moscardini \& Pantano 1996), where we
give a complete description of the method and we show the results of various
numerical tests. Here we only summarize the main characteristics of the code. 

The hydrodynamical part has been developed by using the Eulerian version of the
Piecewise Parabolic Method (PPM; Colella \& Woodward 1984) which ensures at
least second-order (up to the fourth-order, in the case of smooth flows and
small timesteps) accuracy in space and second-order accuracy in time. The high
accuracy of this method allows minimization of errors due to the finite size of
the cells of the grid and leads to a spatial resolution close to the nominal
one (i.e. one grid). In a cosmological framework, the basic PPM technique has
been modified to include the gravitational interaction and the expansion of the
universe. Particular care has been devoted to the calculation of the gas
internal energy and a double formulation of the energy equation has been used
in order to avoid large errors in the computation of thermodynamical quantities
when the kinetic energy is very large compared to the internal one. 

The hydrodynamical part has been coupled to a Particle Mesh (PM) N-body code
(Hockney \& Eastwood 1981) that describes the evolution of the dark component.
The standard PM code has been modified in order to allow non-constant timesteps
equal to those used in the integration of the hydrodynamical equations. This is
obtained by replacing the standard second-order leapfrog method by a
second-order two-step Lax-Wendroff scheme. Densities and forces are computed by
using the cloud-in-cell interpolation scheme. The coupling is obtained by
calculating, by the usual FFT procedure, the gravitational field due to both
components. 

In the simulations presented in this paper we neglected the atomic processes
for radiative cooling since the cooling time of the hot gas produced in
clusters is longer than a Hubble time. Moreover the only considered heating
processes are the adiabatic compression and the entropy generation at shock
fronts. 

For each cosmological model we run one simulation with the initial conditions
given by the same random sequence. The initial redshift, fixed in such a way
that the maximum initial density fluctuation is less than unity, is
approximately $z \approx 20$ for all models. The box-size has been fixed to
$64\ h^{-1}$ Mpc and the number of computational cells is $128^3$. Consequently
the nominal spatial resolution, which our numerical tests have shown to be very
close to the effective one (Gheller et al. 1997), is $0.5\ h^{-1}$ Mpc. In the
analysis of the results we have to be careful to properly evaluate the spurious
effects that the limited box-size and the finite grid resolution can produce on
the results. In fact the grid resolution prevents us from following the
behaviour of matter inside a cell element. This can lead to underestimation of
quantities like density and temperature, with a direct influence on the
evolutionary history of the X-ray clusters. On the other hand, the limited size
of the box has the effect of suppressing the large-scale power, reducing the
possibility of forming very bright X-ray clusters with luminosity larger than
$10^{45}$ erg s$^{-1}$. This kind of object is also likely to be partially
missed in our simulations because it is rare: observational data (e.g. Henry \&
Arnaud 1991; Ebeling et al. 1997) show that in a box of $64\ h^{-1}$ Mpc one
expects to find at most one such object. 

\subsection{Cluster Identification} 

The first step of the data analysis consists of the identification of the X-ray
clusters. We first calculate the emissivity due to thermal bremsstrahlung in a
fully ionized plasma ($X=0.76$, $Y=0.24$) with temperature $T$ (Rybicki \&
Lightman 1979; see also Evrard 1990): 
\ba
\epsilon_{\nu}& = & 6.83 \times 10^{-38}Z^2  n_e n_i T^{-1/2}  \times \nn
& & \bar{g}(h\nu/kT) \ e^{-h\nu/kT}  {\rm erg \ s ^{-1} \ cm^{-3} \ Hz^{-1}},
\ea
where $\bar{g}(h\nu /kT) e^{-h\nu/kT}$ is an average Gaunt factor, $Z$ is the
charge number, $n_i$ and $n_e$ are the ion and the electron density,
respectively. 

The bolometric emissivity, using a unit Gaunt factor, is defined as: 
\ba
\epsilon_{bol}& = & \int_0^\infty d\nu \epsilon_\nu \nn
 & = & 1.42 \times 10^{-27} Z^2 n_i n_e T^{1/2} {\rm erg \ s ^{-1} \ cm^{-3}}. 
\ea
Then the energy radiated within a given energy band $E_1$ - $E_2$ can be
expressed as 
\be
\epsilon_{band}=f_{band}(T) \ \epsilon_{bol}\ ,
\ee
where 
\be
f_{band}(T)=\int_{E_1/kT}^{E_2/kT} d\eta \ \bar{g}(\eta)\ e^{-\eta}\ .
\ee
The band limited X-ray emission $L_x$ from a given volume is computed
integrating the previous expression over the relevant volume. Using the
discretization of the simulation, the X-ray luminosity is 
\be
L_x = 1.25 \times 10^{-27}  m_p^{-2} \sum_i \rho_{_{BM_i}}^2 T_i^{1/2} 
f_{band} (T_i)\ {\rm erg \ s^{-1}}\ ,
\ee
where the sum runs over cells within the volume, $\rho_{_{BM}}$ is the baryon
density and $m_p$ is the proton mass. Since we are interested only in X-ray
emitting regions, it is safe to assume that the gas is completely ionized. We
also set the Gaunt factor $\bar{g}=1.2$: this gives an accuracy of $\sim 20$
per cent for the results. 

At this point, in order to identify the clusters, we select the cells with
$L_x\ge 10^{40}$ erg s$^{-1}$ which are also local maxima in the X-luminosity
field (i.e. their X-luminosity is greater than that of the 26 neighboring
cells). These identify the cluster centres. A cluster is defined as the sum of
the centre plus the 26 surrounding cells. In this way the total volume of a
cluster equals the volume of a sphere of comoving radius $0.93\ h^{-1}$ Mpc, as
appropriate for present observed X-ray clusters. In order to avoid double
counting of the cells, the distance between cores is checked: if two cores are
closer than $2\ h^{-1}$ Mpc the fainter cluster is rejected from the catalogue.
Finally, the cluster luminosity and the mass are calculated as the sum of the
luminosity and density of each of its cells, respectively, while the
temperature is defined as the average over the whole cluster volume. 

\section{Results} 

\subsection{Global properties} 

In Figure 1, we show a snapshot of the results in a slice of $64\times 64\times
0.5\ h^{-3}$ Mpc$^3$ at $z=0$ for each of the four models. The baryonic matter
density field $\varrho_{_{BM}}$, the dark matter density field
$\varrho_{_{DM}}$, the gas temperature $T$ and the X-ray emission $L_x$ are
presented in Figures 1a, 1b, 1c and 1d, respectively. Since the initial
spectrum of fluctuations of all simulations has been generated by using the
same random number sequence, the positions of the final structures are quite
similar and the densities of each component scale approximatively according to
their mean cosmic values. Matter concentrates on filamentary structures and
clusters form at the intersection of several filaments. 

In the low-$\Omega_{BM}$ models, because of the larger power on small scales in
the initial spectrum and the lower background pressure, shocks form earlier and
are stronger than in the high-$\Omega_{BM}$ models. In the latter case, on the
other hand, we can observe a stronger X-ray emission due to the higher baryon
content of these models. 

A close-up at four different redshifts of the most luminous X-ray cluster found
in the BM05 simulation is presented in Figure 2. The slice is $20\times 20
\times 0.5\ h^{-3}$ Mpc$^3$. It is evident the ongoing process of merging and
the final virialized  state characterized by an extended central isothermal
region at high temperature. The cluster tends to a spherical geometry and
deviations from sphericity are due to the cluster memory of its merging history
(see e.g. Tormen 1997). These satellites, however, being faint X-ray sources,
may not be observed in a X-ray map. Similar comments can be repeated for the
other models (not shown here). 

Dark matter structures appear typically less concentrate than the baryonic
counterpart both in clusters and in filaments. In Table 2 we present the rms of
the DM and BM density fields ($\sigma_{_{DM}}$ and $\sigma_{_{BM}}$
respectively) computed on the cell-size scale ($0.5\ h^{-1}$Mpc) and the mean
temperature  $<T>$ (in Kelvin degrees) at redshifts $z=1$ and $z=0$. The
density contrasts of the two components are normalized to the corresponding
mean cosmological values. A detailed analysis of the time evolution of the rms
shows that at very high redshifts, due to absence of pressure forces, DM
collapses faster than the baryonic counterpart. However, in all of our models,
starting from about $z\sim 3$, baryons tend to concentrate more than the dark
component. The absence  of dissipative phenomena produces a spreading of DM
around the minima of the gravitational potential, while BM, which tends to
thermalize, concentrates there. As time goes by, also dark particles fall
toward the centre of the potential well and at the final time the rms of the
two components is quite similar (see Table 2). 

Comparing the different models, we observe that both $\sigma_{_{BM}}$ and
$\sigma_{_{DM}}$ decrease with the increasing of the baryon fraction both at
$z=1$ and $z=0$. This behaviour is mainly related to the characteristics of the
initial spectrum in the different models. Furthermore structures virialize
earlier in low-$\Omega_{BM}$ models and shocks are stronger and the final
temperatures higher than in the high-$\Omega_{BM}$ models, because of the lower
pressure and higher densities present initially in these models. Between $z=1$
and $z=0$ there is a slightly faster evolution of structures in the case of a
high baryon content and this is probably favoured by the lower temperatures
produced in this case. 

\begin{table}
\centering
\caption[]{The global properties of the model simulations. The rms of the dark
matter, $\sigma_{_{DM}}$, the rms of the baryonic matter, $\sigma_{_{BM}}$, and
the mean temperature  $<T>$ (in Kelvin degrees), computed on the cell scale,
are shown at redshift $z=1$ (Columns 2, 3 and 4) and at $z=0$ (Columns 5, 6 and
7).} 
\tabcolsep 4pt
\begin{tabular}{lccccccc} \\ \\ \hline \hline
Models & $\sigma_{_{DM}}$  & $\sigma_{_{BM}}$ & $<T>$ K& 
& $\sigma_{_{DM}}$  & $\sigma_{_{BM}}$ & $<T>$ K
\\ 
& \multicolumn{3}{c} {$z=1$} & & \multicolumn{3}{c} {$z=0$} \\ \hline
BM05  & 3.27 & 3.40 & $4.86\times 10^5$& & 7.31 & 7.34 & $1.04\times 10^6$ \\
BM10  & 2.80 & 3.14 & $3.86\times 10^5$& & 6.97 & 7.00 & $9.24\times 10^5$ \\
BM15  & 2.44 & 2.73 & $2.97\times 10^5$& & 6.05 & 6.27 & $7.68\times 10^5$ \\
BM20  & 1.94 & 2.20 & $2.20\times 10^5$& & 5.33 & 5.45 & $6.35\times 10^5$  \\ 
\hline
\end{tabular}
\end{table}

\subsection{X-ray cluster mass}

Clusters in our simulations are identified through their X-ray emission and
their characteristic properties, like the total mass, the total luminosity and
the mean temperature, are computed by integrating or averaging over a fixed
number of cells, as explained in Section 3.2. The values of these quantities
could then be affected by the choice of the number of cells used for the
calculation. Several tests have shown that our procedure provides good
estimates for the temperature and the luminosity of the cluster. In fact the
temperature is almost uniform over regions greater than those over which we do
our averaging. Moreover the luminosity is proportional to the square of the
baryonic density and then its value depends essentially on the cells with
highest density which represents the centre of the cluster and which are
consequently always included in our integration. The inclusion or missing of
some low-density cells does not affect sensibly our estimates for the
temperature and the luminosity, but could affect the calculation of the cluster
total mass. In summing over a fixed number of cells  we tend to overestimate
the mass of objects which are not as extended as our reference volume (27
cells) and which are also usually characterized by density lower that the mean
cluster values. The opposite is true for large clusters, although in this case
we have verified that, for our choice of the reference volume, the error
introduced in the estimation of the mass is less severe and it is at most a
factor two. Therefore we have restricted our analysis to clusters with mass
greater than $10^{14} M_\odot$ (for $h=0.5$). This lower limit has been fixed
in order to avoid the inclusion of too low-mass objects which can be affected
by a large error in the mass estimate and whose properties differ considerably
from those of a typical galaxy cluster (e.g., their mean density is much lower
than that determined from observations). Furthermore their spatial distribution
could be affected by the X-luminosity selection criterion used to build our
cluster catalogue. 

In Figure 3 we present, for the four models, the number $N_M$ of clusters with
mass greater than $10^{14} M_\odot$ found in the whole simulation box at
various redshifts. The number of clusters decreases with increasing
$\Omega_{BM}$ at any redshift. This is a consequence of the amount of power on
small scales, which decreases as we increase the baryonic fraction. In the
high-$\Omega_{BM}$ cases the collapse of massive objects occurs later and,
then, if we select clusters  by their mass, the models with high baryonic
fraction presents less clusters than low-$\Omega_{BM}$ models. The behaviour of
the mass function depends on the distribution of the total (dark plus baryonic)
density and then it is not the preferred quantity for discriminating between
our models. Other quantities, like the X-ray luminosity, that depends directly
on the BM density, appear to be a more useful quantity for the problem studied
in the present paper. 

Notice that the estimated $N_M$ roughly agrees with the observational data for
all the models. This is an expected result, because of the spectra
normalizations. In fact it is known that the present abundance of galaxy
clusters requires, in the framework of the critical-density models and almost
independently of the shape of the primordial spectrum, a normalization
$\sigma_8\approx 0.6$, with a quite large uncertainty (e.g. White, Efstathiou
\& Frenk 1993; Viana \& Liddle 1996; Eke, Cole \& Frenk 1996). 

\subsection{X-ray cluster luminosity} 

In Figure 4 we present the number of clusters with luminosity greater than
$L_x=10^{42}$ and $10^{43}$ erg\ s$^{-1}$ ($N_{42}$ and $N_{43}$,
respectively). This luminosity-based selection criterion leads to results
opposite to those obtained by selecting clusters by their masses. In fact at
any epoch the number of clusters increases with increasing $\Omega_{BM}$.
Furthermore the evolution of $N_{42}$ and $N_{43}$ with redshift is not
monotonic: the number of clusters tends to grow until a turn-around redshift
after which it starts to decrease. This behaviour is common to all the models
and for both the adopted minimum luminosity. The only exception is $N_{43}$ in
the case of BM05, which grows continuously with time, but, because of the small
number, this might be not statistically significant. The turn-around is due to
the balance between the mechanisms driving the cluster evolution, and it is an
indication of the epoch when the merging processes of different structures
start to dominate over the gravitational collapse of each single object. In
fact the merging leads to larger but smoother structures. Since the X-ray
emission is proportional to the square of the baryonic density, lower
luminosities are expected. The effect of the merging processes is also shown by
the simultaneous decrease in the number of fainter clusters (not reported in
the figure). The turn-around redshift becomes lower with increasing the baryon
fraction ranging from $z=0.7$ for BM05 to $z=0.2-0.5$ for BM20. This is due to
the delayed evolution of the structures in high-$\Omega_{BM}$ models. 

This behaviour is confirmed by the X-ray emissivity per unit comoving volume
due to both the gas in its entirety, $j_{gas}$, and the clusters, $j_{cl}$ (see
Table 3). For each model the two quantities evolve in a parallel way,
indicating that clusters emit roughly a constant large fraction of the total
X-ray radiation. Similar results have been obtained, for the standard CDM
model, also by Kang et al. (1994a) and Bryan et al. (1994a). 
  
\begin{table}
\centering
\caption[]{The X-ray emissivity (in units of $10^{40}$ ergs s$^{-1}$ $h^3$
Mpc$^{-3}$) for the gas ($j_{gas}$) and for the clusters ($j_{cl}$) 
at various redshifts for the different models.}
\tabcolsep 4pt
\begin{tabular}{lccccccccc} \\ \\ \hline \hline
 & \multicolumn{2}{c} {BM05} & \multicolumn{2}{c} {BM10} 
& \multicolumn{2}{c} {BM15} & \multicolumn{2}{c} {BM20} \\
\hline
& $j_{gas}$ & $j_{cl}$ & $j_{gas}$ & $j_{cl}$ & $j_{gas}$  & $j_{cl}$  
& $j_{gas}$  & $j_{cl}$\\  
\hline
$z=1$   & 0.17 & 0.12 & 0.49 & 0.39 & 0.66 & 0.49 & 0.73 & 0.55\\ 
$z=0.7$ & 0.21 & 0.16 & 0.54 & 0.43 & 0.79 & 0.62 & 0.98 & 0.79\\ 
$z=0.5$ & 0.20 & 0.15 & 0.63 & 0.46 & 0.88 & 0.73 & 1.53 & 0.86\\
$z=0.2$ & 0.19 & 0.14 & 0.51 & 0.39 & 0.87 & 0.73 & 1.29 & 1.12\\
$z=0$   & 0.16 & 0.13 & 0.59 & 0.50 & 0.96 & 0.81 & 1.05 & 0.85\\
\hline
\end{tabular}
\end{table}

The fundamental differences between the cluster abundances in mass ($N_M$) and
in X-ray luminosity ($N_{42}$ and $N_{43}$) must be kept in mind when the
results are compared to the observations. In Figure 5 we present the luminosity
function for the four models computed at five different redshifts, integrated
over the whole range of frequencies. Very bright clusters with luminosity
greater than $10^{45}$ erg\ s$^{-1}$ are missing in our simulations. The lack
of such clusters can be related to two different effects. Firstly, the size of
our computational box limits the amount of large-scale power which we can
follow in the simulation, and consequently the maximum temperature that can be
produced. Secondly, the grid resolution is likely to underestimate the highest
density peaks where brightest clusters are expected to form. 

The luminosity functions have been fitted by using a two-parameter function: 
\be
n(L) dL\ =\ n_0L^{- \alpha}dL\ ,
\label{eq:two}
\ee
where $n(L)dL$ is the comoving density of clusters with luminosity between $L$
and $L+dL$, $n_0$ is in units of $10^{-6}\ h^3$Mpc$^{-3}$ and $L$ is in units
of $10^{44}$ erg\ s$^{-1}$. The results of the fits are presented in Table 4.
Notice that for our data a two-parameter fit is more appropriate than the usual
Schechter function, as in our results the expected bend at high $L$ is not
present, for the reasons given above. 

The values of the slope $\alpha$ cannot discriminate between the four models,
as the differences are in general within the 1-$\sigma$ errorbars. The
parameter $\alpha$ shows for all the models a slight negative evolution with
redshift from $z=0.5$ to $z=0$; this behaviour is thought to be the effect of
the ongoing processes of gravitational collapse, that produces more
concentrated and bright structures, and merging, that leads to the formation of
larger and more massive objects at the expenses of the smaller ones. For the
normalizations $n_0$, which is the comoving number density of objects with
luminosity equal to $10^{44}$ erg\ s$^{-1}$, we can make considerations similar
to those previously made for $N_{42}$ and $N_{43}$. 

\begin{table*}
\centering
\caption[]{The parameters of the fits of the X-ray cluster bolometric
luminosity function $n(L) =n_0L^{-\alpha}$ at various redshifts for the
different models.} 
\tabcolsep 4pt
\begin{tabular}{lccccccccc} \\ \\ \hline \hline
 & \multicolumn{2}{c} {BM05} & \multicolumn{2}{c} {BM10} 
& \multicolumn{2}{c} {BM15} & \multicolumn{2}{c} {BM20} \\
\hline
& $\alpha$ & $n_0$ & $\alpha$ & $n_0$ & $\alpha$  & $n_0$  
& $\alpha$  & $n_0$\\  
\hline

$z=1$   & 1.82$\pm$ 0.09 & ~6.73$\pm$ 0.06 & 1.72$\pm$ 0.08 & 18.95$\pm$ 0.23 & 
1.59$\pm$ 0.07 & 31.66$\pm$ 0.38 & 1.74$\pm$ 0.06  & 18.56$\pm$ 0.15 \\ 

$z=0.7$ & 1.73$\pm$ 0.15 & ~9.72$\pm$ 0.18 & 1.72$\pm$ 0.09 & 21.55$\pm$ 0.30 & 
1.71$\pm$ 0.05 & 26.60$\pm$ 0.24 & 1.73$\pm$ 0.07  & 28.44$\pm$ 0.33 \\ 

$z=0.5$ & 1.74$\pm$ 0.20 & ~8.42$\pm$ 0.20 & 1.75$\pm$ 0.16 & 19.46$\pm$ 0.47 & 
1.76$\pm$ 0.09 & 25.66$\pm$ 0.36 & 1.69$\pm$ 0.08  & 33.20$\pm$ 0.46 \\

$z=0.2$ & 1.61$\pm$ 0.07 & 12.50$\pm$ 0.12 & 1.68$\pm$ 0.04 & 20.97$\pm$ 0.12 & 
1.69$\pm$ 0.05 & 29.12$\pm$ 0.24 & 1.69$\pm$ 0.11  & 35.59$\pm$ 0.72 \\

$z=0$   & 1.41$\pm$ 0.11 & 23.11$\pm$ 0.48 & 1.67$\pm$ 0.11 & 18.68$\pm$ 0.31 & 
1.62$\pm$ 0.05 & 31.69$\pm$ 0.25 & 1.66$\pm$ 0.03  & 33.95$\pm$ 0.15 \\
\hline
\end{tabular}
\end{table*}

The hatched region in the figure shows the observational data (with 1-$\sigma$
errorbars) of Ebeling et al. (1997) which refer to the ROSAT Brightest Cluster
sample containing 199 objects with redshift $z\le 0.3$. The observational
curves have been fitted by the authors by using a three-parameter function: 
\be
n(L)\ =\ A\exp (-L/L^*)L^{-\alpha}\ ,
\label{eq:three}
\ee
where $A$ is in units of $10^{-7}$Mpc$^{-3}$($10^{44}$erg\ s$^{-1}$)$
^{\alpha-1}$ and $L^*$ is in units of $10^{44}$erg\ s$^{-1}$. The values of the
fitting parameters are $A=6.41^{+0.70}_{-0.61}$, $L^*= 37.2^{+16.4}_{-3.8}$ and
$\alpha=1.84^{+0.09}_{-0.04}$. 

We can compare these data with the results of the simulations at low redshifts.
The models with high baryonic content (BM15 and BM20) have a luminosity
function which is significatively too high with respect to the observations. On
the contrary the BM05 model and (much more marginally) BM10 are in better
agreement with the data. 

Similar conclusion can be obtained if we consider the luminosities in the
energy band [0.5--2] keV. The fitting parameters for our simulations are
reported in Table 5, while the comparison with two different observational
datasets is shown in Figure 6. The vertically hatched region refers again to
the Ebeling et al. (1997) sample, whose luminosity function has been fitted by
a three-parameter relation (\ref{eq:three}) with $A=3.32^{+0.36}_{-0.33}$,
$L^*= 5.70^{+1.29}_{-0.93}$ and $\alpha=1.85^{+0.09}_{-0.09}$. The horizontally
hatched region shows instead the results obtained by De Grandi (1996) using a
complete flux-limited ROSAT sample selected from the ESOKP redshift survey: in
this case the fitting parameters are $A=4.51$, $L^*=2.63^{+0.87}_{-0.58}$ and
$\alpha=1.32^{+0.21}_{-0.23}$. These two determinations of the luminosity
functions are in good agreement for luminosities larger than $\approx 2 \times
10^{43}$ erg s$^{-1}$, while for smaller $L_x$ the De Grandi (1996) results are
approximately a factor 3 smaller than the Ebeling et al. (1997) ones,
increasing the discrepancies between the observations and the model predictions
of the models with high $\Omega_{BM}$. 

In this energy band there is a further luminosity function determined by Burns
et al. (1996), always by using images from the ROSAT all-sky survey. Because of
the large errorbars, it completely overlaps both the previous results and for
clarity we prefer do not show it in Figure 6. However, since in this dataset
also nearby poor clusters have been considered, this result allows to extend
the previous considerations also to smaller X-ray luminosities (less than
$10^{42}$ erg s$^{-1}$), not included in the other datasets. 

\begin{table*}
\centering
\caption[]{The parameters of the fits of the X-ray cluster luminosity function
$n(L) =n_0L^{-\alpha}$ computed in the [0.5--2] keV band at various redshifts
for the different models.} 
\tabcolsep 4pt
\begin{tabular}{lccccccccc} \\ \\ \hline \hline
 & \multicolumn{2}{c} {BM05} & \multicolumn{2}{c} {BM10} 
& \multicolumn{2}{c} {BM15} & \multicolumn{2}{c} {BM20} \\
\hline
& $\alpha$ & $n_0$ & $\alpha$ & $n_0$ & $\alpha$  & $n_0$  
& $\alpha$  & $n_0$\\  
\hline
$z=1$ & 1.47$\pm$ 0.11  & 11.73$\pm$ 0.17 & 1.43$\pm$ 0.07 & 19.26$\pm$ 0.24 &
1.39$\pm$ 0.10 & 19.19$\pm$ 0.31 & 1.40$\pm$ 0.11   & 10.17$\pm$ 0.13 \\
$z=0.7$ & 1.58$\pm$ 0.40  & ~8.40$\pm$ 0.49 & 1.55$\pm$ 0.11 & 15.73$\pm$ 0.26 &
1.58$\pm$ 0.11 & 12.90$\pm$ 0.20 & 1.45$\pm$ 0.12   & 18.25$\pm$ 0.32 \\
$z=0.5$ & 1.81$\pm$ 0.16  & ~5.52$\pm$ 0.04 & 1.58$\pm$ 0.12& 15.09$\pm$ 0.28 &
1.52$\pm$ 0.07 & 22.51$\pm$ 0.30 & 1.57$\pm$ 0.10   & 28.56$\pm$ 0.48 \\
$z=0.2$ & 1.44$\pm$ 0.11  & 13.71$\pm$ 0.27 & 1.55$\pm$ 0.06 & 15.87$\pm$ 0.17 &
1.67$\pm$ 0.08 & 13.63$\pm$ 0.15 & 1.55$\pm$ 0.11   & 21.31$\pm$ 0.52 \\
$z=0$   & 1.36$\pm$ 0.17  & 18.05$\pm$ 0.60 & 1.63$\pm$ 0.17 & 11.75$\pm$ 0.28 &
1.61$\pm$ 0.06 & 11.75$\pm$ 0.28 & 1.57$\pm$ 0.09   & 19.28$\pm$ 0.28  \\
\hline
\end{tabular}
\end{table*}

\subsection{X--ray cluster temperatures}

In Figure 7 we show the redshift evolution of the distribution of the cluster
mean temperature for the four models. The temperatures have been calculated as
emission-weighted averages because this is the quantity which is also usually
estimated from the observations. The absence of clusters with temperatures
above 4 keV is mainly related to the limited size of the box and their
rareness. These model predictions can be compared with the observations. In the
figure the hatched region refers to the temperature distribution obtained by
Henry \& Arnaud (1991) from a set of local ($z\approx 0$) clusters: 
\be
n(T)\ =\ (1.8^{+0.8}_{-0.5}\times 10^{-3}\ 
h^3 \ {\rm Mpc}^{-3} \ {\rm keV}^{-1})\ T^{-4.7\pm 0.5}, 
\ee 
where the uncertainties are 1-$\sigma$ errorbars and $T$ is expressed in keV. 

All the models are in quite good agreement with observations in the overlapping
range. Temperature is in fact less sensitive than luminosity to the details of
the density distribution and it is related to the maximum wavelength $\lambda$
of non-linear waves. In fact the post-shock temperature is of the order of
$T\propto (H\lambda)^2$, where $H$ is the Hubble constant. Low-$\Omega_{BM}$
models have a higher normalization of the primordial spectrum and longer
wavelengths can reach the non--linear regime at the final time producing higher
values of the temperature. This phenomenon  is likely to be strengthened by the
different mean background pressure of the various realizations, which is lower
in low-$\Omega_{BM}$ models. Both effects could explain the higher number of
objects with temperature larger than about 1 keV found with decreasing
$\Omega_{BM}$. 

We have also analysed the redshift evolution of the temperature distributions
in our simulations. We found that between $z=1$ and $z=0$ such distributions
are almost constant for BM05 and BM10 models, with a slow increase in the
number of high-temperature objects. In these models, by $z\sim 1$ the regions
heated up by shocks at almost a uniform temperature are larger than the
integration volume of our cluster identification method. In general, from that
moment, the temperature of these regions increases because of adiabatic
compression and merging processes and this explains the rise in the number of
high-temperature clusters towards $z=0$. The models BM15 and BM20, instead,
evolve rapidly between $z=1$ and $z=0.5$, showing in particular a strong growth
in the number density of objects with temperature greater than about 0.5 keV.
This corresponds to the later formation and propagation of the shocks in these
models. In fact after $z=0.5$ the situation becomes similar to that of the
low-$\Omega_{BM}$ models, and the temperature distributions show little further
evolution. 
 
\subsection{Luminosity-Temperature relation} 

In Figure 8 we present the distribution of emission-weighted temperature of the
clusters as a function of the X-ray bolometric luminosity at three different
redshifts: $z=1$ (crosses), $z=0.5$ (open circles) and $z=0$ (filled circles).
For all the models there is a similar trend in the luminosity-temperature
relation, even though the total number of objects grows with the baryon
fraction. We observe that a given temperature corresponds to higher
luminosities in high-$\Omega_{BM}$ models. This is related to the higher
baryonic densities present in these cases. 

The luminosity-temperature distributions have been fitted by using a power-law
relation of the form $T\ =\ 10^{b}L_x^{\eta}$, where $T$ is in keV and $L_x$ is
in units of $10^{40}$erg\ s$^{-1}$. The results, reported in Table 6, show that
for all the models there is a little evolution with time, especially between
$z=0.5$ and $z=0$, where present-day observational data are available. In
particular the parameter $\eta$ is always inside the 1-$\sigma$ errorbar range,
even if the models with smaller baryonic content tend to have a steeper slope.
The normalization $b$, instead, slightly decreases with increasing redshift
and/or $\Omega_{BM}$. This result is in qualitative agreement with the more
recent observational analysis, which found no evolution of the
temperature-luminosity distribution, at least for $z<0.5$ (Mushotzky \& Scharf
1997). 

\begin{table*}
\centering
\caption[]{The parameters of the fits of the Luminosity-Temperature relation
$T\ =\ 10^{b}L_x^{\eta}$, computed at various redshifts for the different
models.} 
\tabcolsep 4pt
\begin{tabular}{lccccccccc} \\ \\ \hline \hline
 & \multicolumn{2}{c} {BM05} & \multicolumn{2}{c} {BM10} 
& \multicolumn{2}{c} {BM15} & \multicolumn{2}{c} {BM20} \\
\hline
& $\eta$ & $b$ & $\eta$ & $b$ & $\eta$  & $b$  
& $\eta$  & $b$\\  
\hline
$z=1$   & 0.41$\pm$0.04 & -1.22$\pm$0.09 & 0.34$\pm$0.03 & -1.31$\pm$0.07& 
0.30$\pm$0.04 & -1.38$\pm$0.10 & 0.31$\pm$0.04 & -1.52$\pm$0.11\\ 

$z=0.7$ & 0.40$\pm$0.03 & -1.10$\pm$0.07 & 0.33$\pm$0.03 & -1.20$\pm$0.09 & 
0.31$\pm$0.04 & -1.32$\pm$0.09 & 0.30$\pm$0.03 & -1.42$\pm$0.08 \\ 

$z=0.5$ & 0.42$\pm$0.03 & -1.06$\pm$0.07 & 0.35$\pm$0.03 & -1.17$\pm$0.09 & 
0.33$\pm$0.03 & -1.28$\pm$0.07 & 0.30$\pm$0.03 & -1.35$\pm$0.07 \\

$z=0.2$ & 0.42$\pm$0.02 & -0.99$\pm$0.06 & 0.36$\pm$0.02 & -1.11$\pm$0.06 & 
0.33$\pm$0.02 & -1.17$\pm$0.06 & 0.31$\pm$0.03 & -1.25$\pm$0.08 \\

$z=0$   & 0.40$\pm$0.02 & -0.90$\pm$0.06 & 0.37$\pm$0.02 & -1.05$\pm$0.05 & 
0.35$\pm$0.02 & -1.14$\pm$0.05 & 0.32$\pm$0.02 & -1.20$\pm$0.06 \\
\hline
\end{tabular}
\end{table*}

Our results can be directly compared to the observational data. The hatched
region in Figure 8 shows the $L_x$-$T$ relation (always with 1-$\sigma$
errorbars) obtained by Henry \& Arnaud (1991) for clusters with luminosity
larger than $10^{44}$ erg s$^{-1}$. Even if the statistics are poor, the most
luminous clusters for all the models are in good agreement with these data. The
dotted line, instead, shows the fit from the combined sample of David et al.
(1993), which contains clusters with lower luminosities ($L_x > 10^{42}$ erg
s$^{-1}$). In this case the dispersion of the data around the fit (not
explicitly reported in the original paper) is of the same order of that showed
for the Henry \& Arnaud (1991) results. The models with high baryonic fraction,
even if they reproduce well the slope of the relation, have a lower
normalization: at the same luminosity, the temperature is at least a factor 3
smaller than for the observations. The agreement is better for BM05 and BM10
models which have a steeper (but still consistent) slope. 

Notice that it is not possible to present a comparison with the relation
obtained by Mushotzky \& Scharf (1997) because it refers to clusters with
luminosities higher than those reached in our simulations. 

\section {Conclusions} 

In this work we have studied the evolution and the properties of X-ray clusters
of galaxies in four different critical-density CDM-like models, in which the
baryon fraction has been varied from $\Omega_{BM}=0.05$ to $\Omega_{BM}=0.20$.
Models with a baryonic content larger than the predictions of the standard
nucleosynthesis have been firstly considered by White et al. (1996) who found
that they are in good agreement with a large set of observational data when
coupled with a small tilt in the primordial spectrum ($n\approx 0.8$). Our
results have proved to be useful in order to discriminate between the various
models and to decide which of these models, if any, is compatible with
observations. 

The mass function, the luminosity function, and the luminosity-temperature
relation are the quantities that gave the most important hints on the
properties of the models. The behaviour of these quantities is determined by
the dynamical evolution of the clusters. This is driven from the balance of two
phenomena: the gravitational collapse of single objects and the merging of
different structures. In the early stages of the evolution, the first effect
tends to dominate, and the X-ray emission grows rapidly. The collapse of the
baryonic matter is stopped by the formation of the shock. This rises strongly
the pressure of the matter that finally is able to contrast the gravitational
infall. Then the smaller virialized objects start to merge together, forming
larger structures characterized by smoother density fields and hence by lower
X-ray emission. 

The mass function presents the expected behaviour, with the cluster number
density that, at $z=0$ decreases with increasing baryon fraction. This is
mainly due to the different amount of power on small scales in the initial
spectra of the density fluctuations. This result is obtained by identifying
clusters only by their mass, without considering their luminosity. On the other
hand, when clusters are selected by their total X-ray luminosity, the opposite
trend is found: high-$\Omega_{BM}$ models have the higher number density of
X-ray clusters. This is essentially related to the presence of more baryons and
so to the formation of higher baryonic density peaks that then leads to much
higher X-luminosity, this quantity depending on the square of the baryonic
density itself. The X-ray emission depends also on the cluster temperatures but
these are roughly the same for all the models. 

Another difference is shown by the time evolution: while the number of clusters
with large mass is a growing function of time for all the models, the abundance
of luminous X-ray clusters starts to decrease at a some redshift, which is
dependent on the cosmological model, being lower for high-$\Omega_{BM}$ models.

We compare the predictions of the four different cosmological models to a
series of observational results, mainly referring to local ($z\approx 0$)
datasets. By analysing the luminosity function and (more marginally) the
luminosity-temperature relation we can conclude that the models with low
baryonic content ($\Omega_{BM}\mincir 0.05$) are in better agreement with the
data, while models with 15 or 20 per cent are well outside the 1-$\sigma$
errorbars. On the contrary the study of the cluster temperature distribution
cannot distinguish between the various models. Our result is even more
stringent if we observe that our resolution tends to underestimate the
luminosities; consequently the differences of the high-$\Omega_{BM}$ models
with observations would be even higher. Therefore the X-ray properties seem to
exclude that the increase of the baryonic content can help to reconcile, in the
framework of critical-density models, the cold dark matter scenario with the
observations. 

In order to investigate if the differences found in the results for the various
cosmological models are due to the different power spectrum normalization (i.e.
$\sigma_8$), we ran a second  simulation for the BM15 model, but adopting in
this case the same $\sigma_8$ value of the BM05 model which we found previously
to be the model in better agreement with the X-ray cluster data. As expected,
in this new simulation with a higher normalization (hereafter BM15$_{hn}$) the
number $N_M$ of clusters with mass greater than $10^{14} M_\odot$ is larger
than in the BM15 model: we found $N_M=28$ and 72 at $z=0.5$ and $z=0$,
respectively. These results are comparable to those obtained for the BM05
model, confirming that the mass function depends mainly on the spectrum
normalization. A similar comment can be done for the rms of the dark and
baryonic matter and for the mean temperature computed on the cell scale. We
also observe a larger formation of X--ray clusters in the BM15$_{hn}$ model
than in the BM15 model: at $z=0$ we found $N_{42}=136$ and $N_{43}=38$. The
resulting luminosity functions, both bolometric and in the energy band [0.5--2]
keV, are slightly higher than those of the BM15 model, increasing the
difference with the observational datasets. A stronger discrepancy with the
data is also obtained when the luminosity--temperature relation is considered.
All these results support the idea that an increase of the primordial baryonic
fraction with respect to the standard values (approximately 5 per cent)
produces X-ray clusters with properties in disagreement with the available
data. Of course this is obtained by assuming a particular set of parameters,
namely $n=0.8$ and $h=0.5$ and it would be interesting to know if the
conclusion survives when these parameters are varied inside the range still
allowing an acceptable fit to other data (see White et al. 1996). 

Finally, another important feature which we found in our simulations is the
very little redshift evolution of X--ray cluster properties. In all of our
models, the luminosity function, the temperature distribution and the
luminosity-temperature relation for the simulated clusters are almost constant,
particularly between $z=0.5$ and $z=0$, in good agreement with that which seems
to emerge also from very recent observational data. 

\section* {Acknowledgments}  

We would like to thank A. Heavens, F. Lucchin, S. Matarrese and G. Tormen for a
critical reading of the manuscript. We are grateful also to A. Cavaliere for
useful discussions. Finally, the referee Andrew Liddle is acknowledged for
valuable comments and criticisms which improved the presentation of the
results. This work was partially supported by Italian MURST.

\newpage
\begin{figure*} 
\centerline{\psfig{file=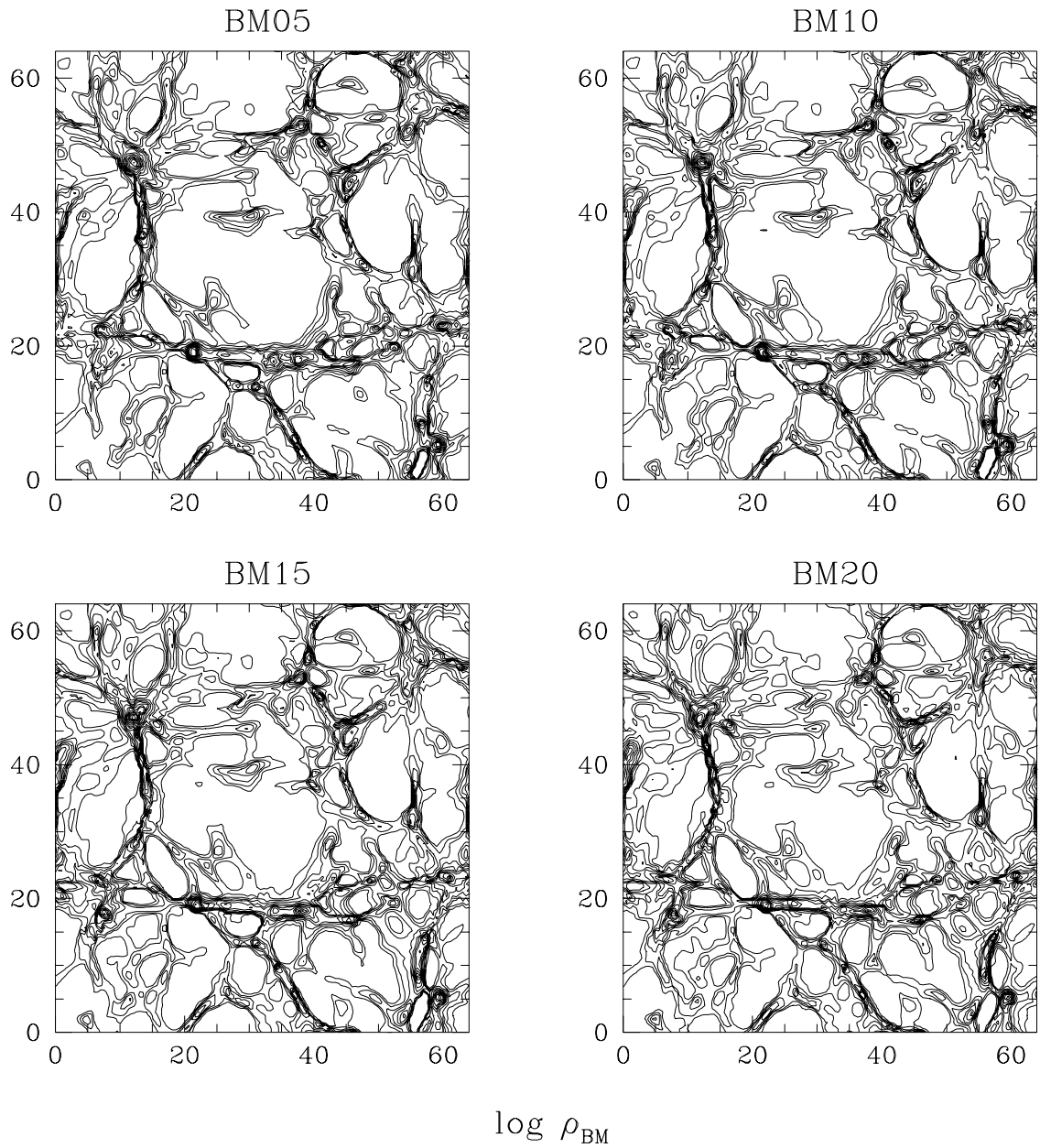,width=17.0cm,height=17.0cm}} 
\caption{
{\bf a.} 
The contour plots for the baryonic density $\varrho_{_{BM}}$ in a slice of
$64\times 64\times 0.5\ h^{-3}$ Mpc$^3$ at $z=0$ for the four different models:
BM05 (top left), BM10 (top right), BM15 (bottom left), BM20 (bottom right). The
baryonic density is normalized to its mean density and the contour levels
correspond to $10^{(i-3)/4}$, where $i=1,2,...,15$ 
}
\end{figure*} 

\newpage
\begin{figure*} 
\centerline{\psfig{file=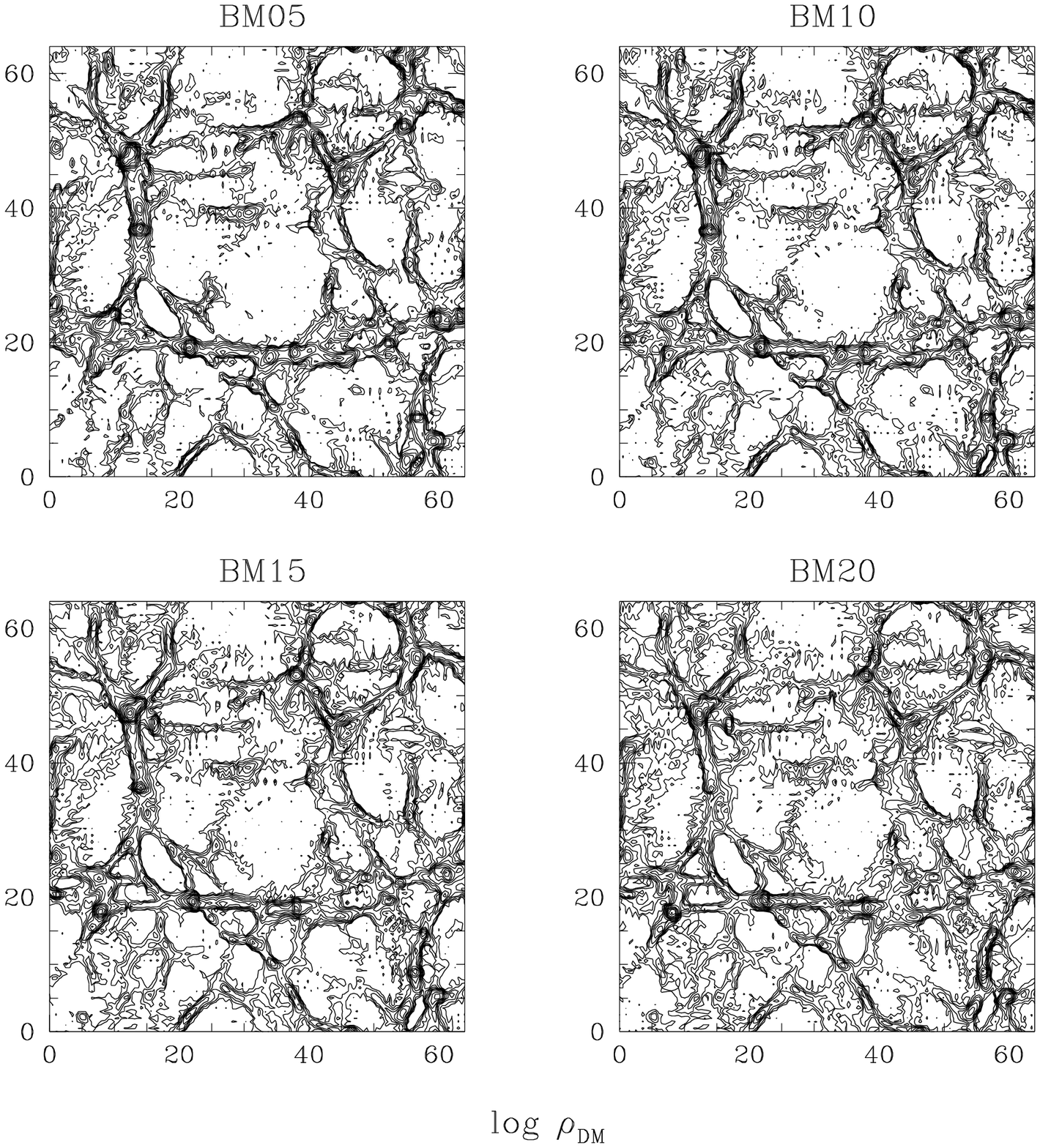,width=17.0cm,height=17.0cm}} 
\caption*{
{\bf b.} 
The same as Figure 1a but for the dark matter density $\varrho_{_{DM}}$ which
is normalized to its mean density. The contour levels correspond to
$10^{(i-3)/4}$, where $i=1,2,...,14$ 
}
\end{figure*} 

\newpage
\begin{figure*} 
\centerline{\psfig{file=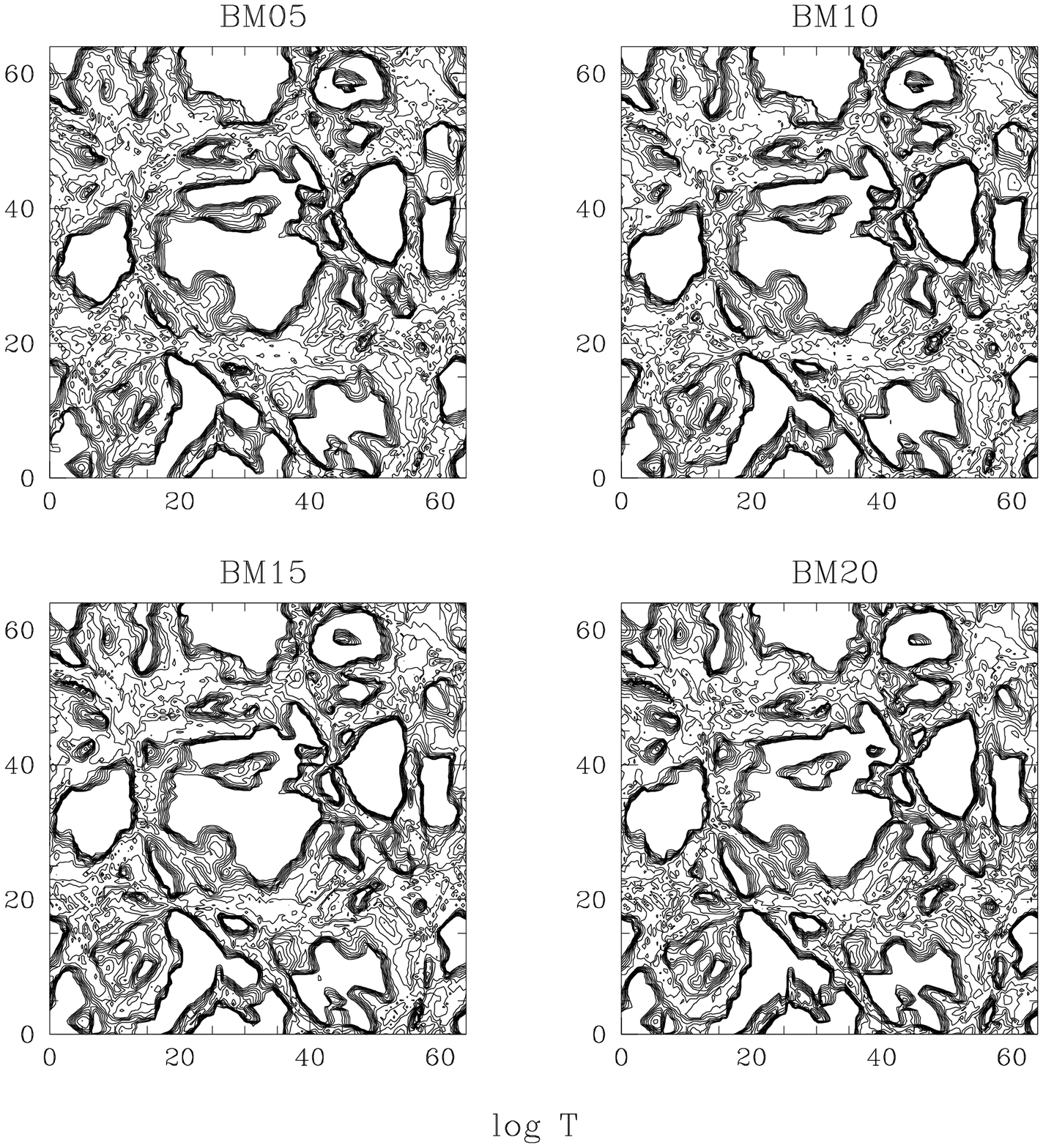,width=17.0cm,height=17.0cm}} 
\caption*{
{\bf c.} 
The same as Figure 1a but for the temperature $T$ which is in units of Kelvin
degrees. The contour levels correspond to $10^{2i/3}$, where $i=1,2,...,12$ 
}
\end{figure*} 

\newpage
\begin{figure*} 
\centerline{\psfig{file=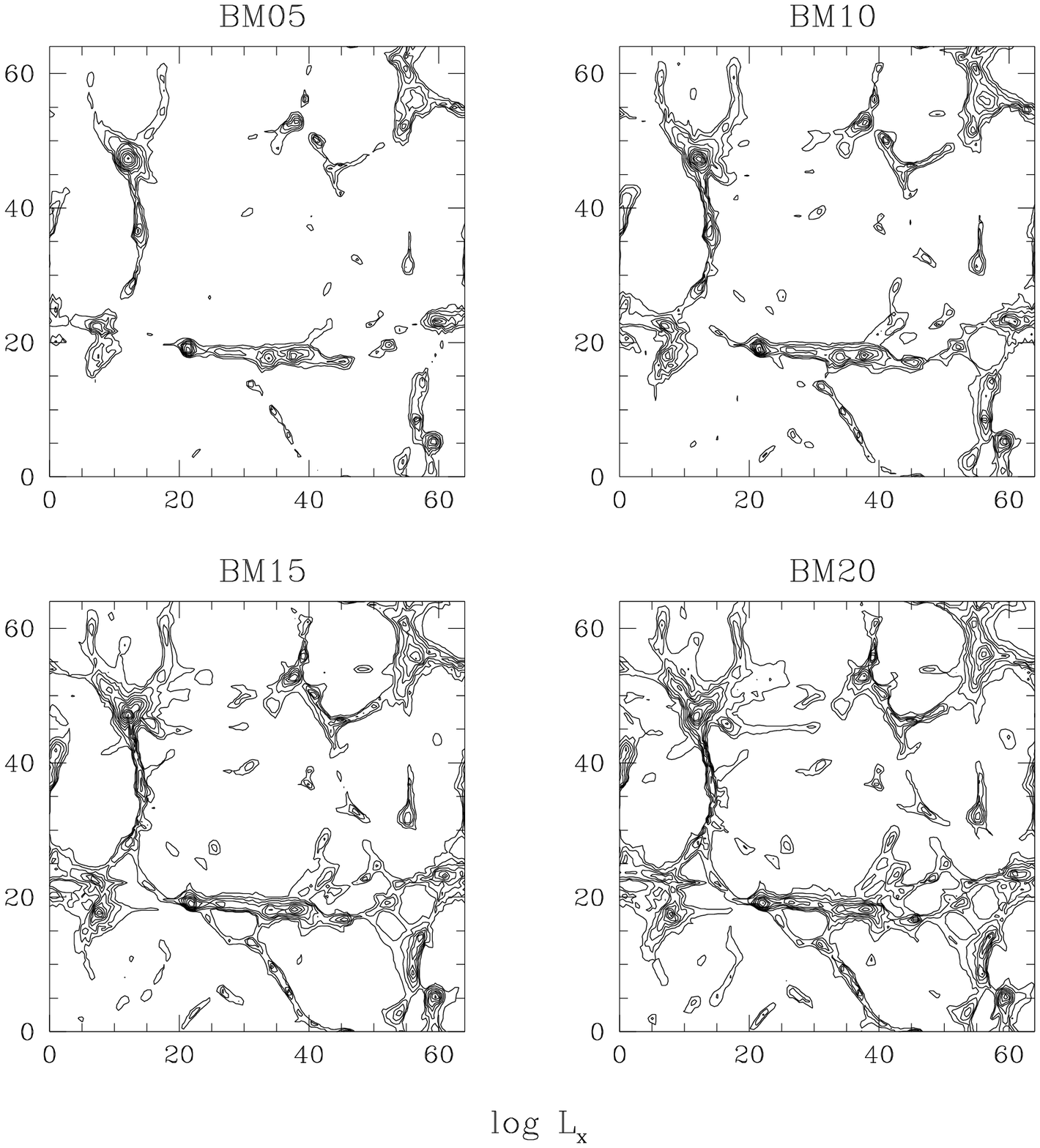,width=17.0cm,height=17.0cm}} 
\caption*{
{\bf d.} 
The same as Figure 1a but for the X-ray luminosity $L_x$ which is in units of
$10^{36}$ erg s$^{-1}$. The contour levels correspond to $10^{2i/3}$, where
$i=1,2,...,11$ 
}
\end{figure*} 

\newpage
\begin{figure*} 
\centerline{\psfig{file=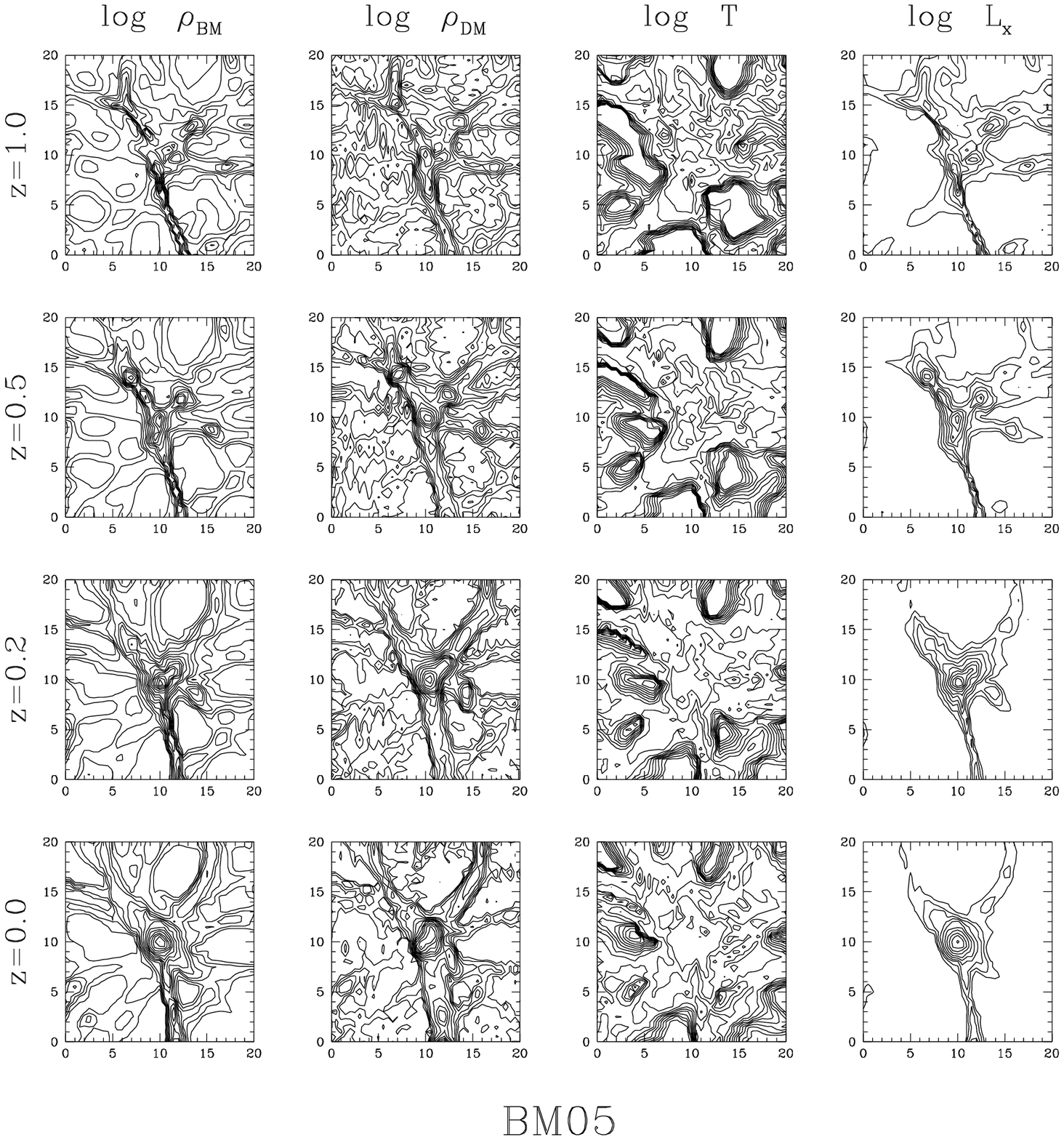,width=17.0cm,height=17.0cm}} 
\caption{
The contour plots for the baryonic density $\varrho_{_{BM}}$ (first column),
the dark matter density $\varrho_{_{DM}}$ (second column), the temperature $T$
(third column) and the X-ray luminosity $L_x$ (last column) in a slice of
$20\times 20 \times 0.5\ h^{-3}$ Mpc$^3$ around the most luminous cluster in
the BM05 simulation. The different rows show the redshift evolution: $z=1$,
$z=0.5$, $z=0.2$ and $z=0$ from the top to the bottom. The density of each
component is normalized to its mean density while the temperature and the
luminosity are in units of Kelvin degrees and $10^{36}$ erg s$^{-1}$,
respectively. The density contour levels correspond to $10^{(i-3)/4}$ while the
temperature and the luminosity levels are $10^{2i/3}$, where $i=1,2,...$ 
}
\end{figure*} 

\newpage
\begin{figure*} 
\centerline{\psfig{file=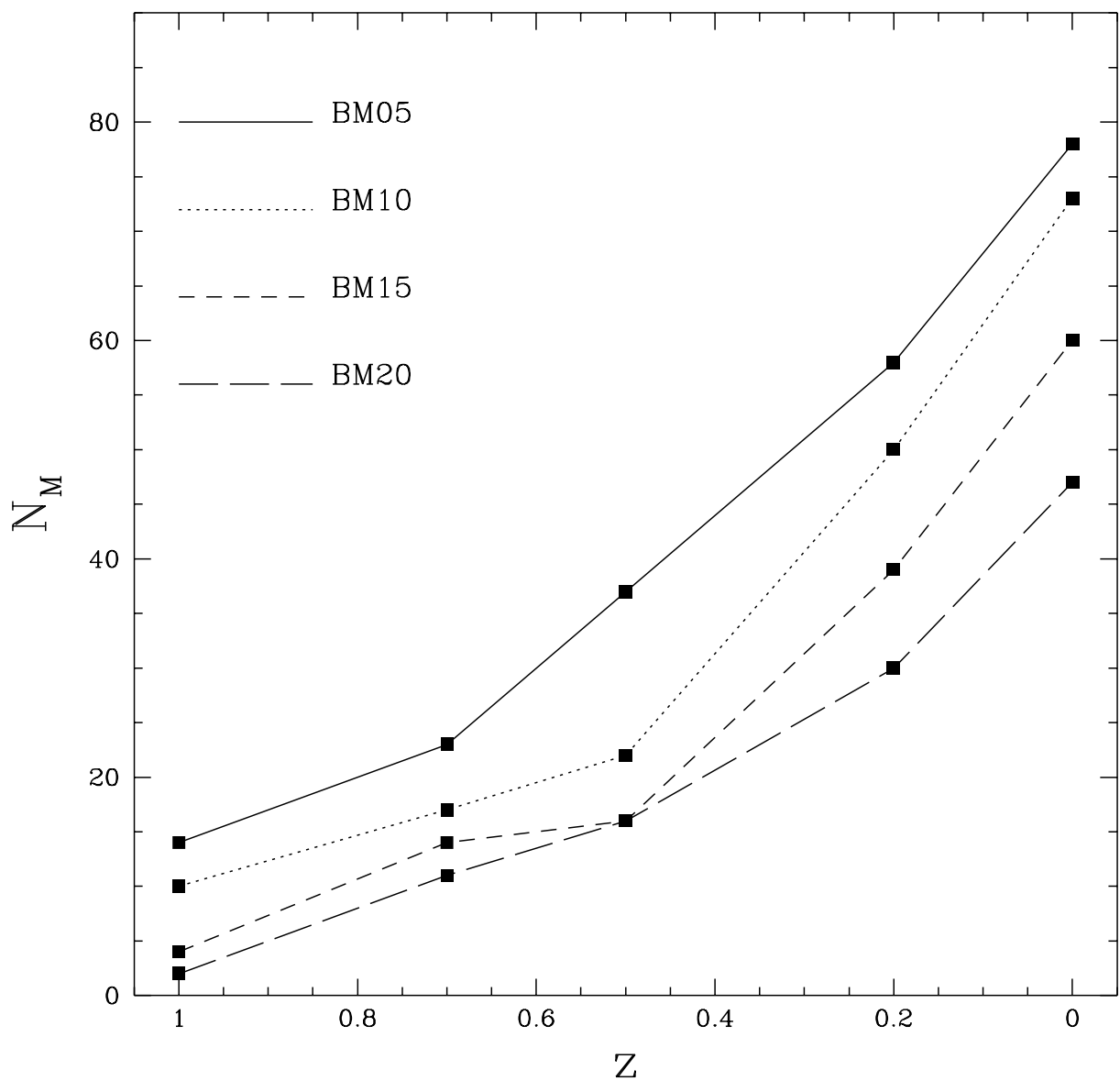,width=17.0cm,height=17.0cm}} 
\caption{
The number of clusters $N_M$ with mass $\ge 10^{14} M_\odot$ as a function of
the redshift $z$ for the different models: BM05 (solid line), BM10 (dotted
line), BM15 (short-dashed line) and BM20 (long-dashed line). 
}
\end{figure*} 

\newpage
\begin{figure*} 
\centerline{\psfig{file=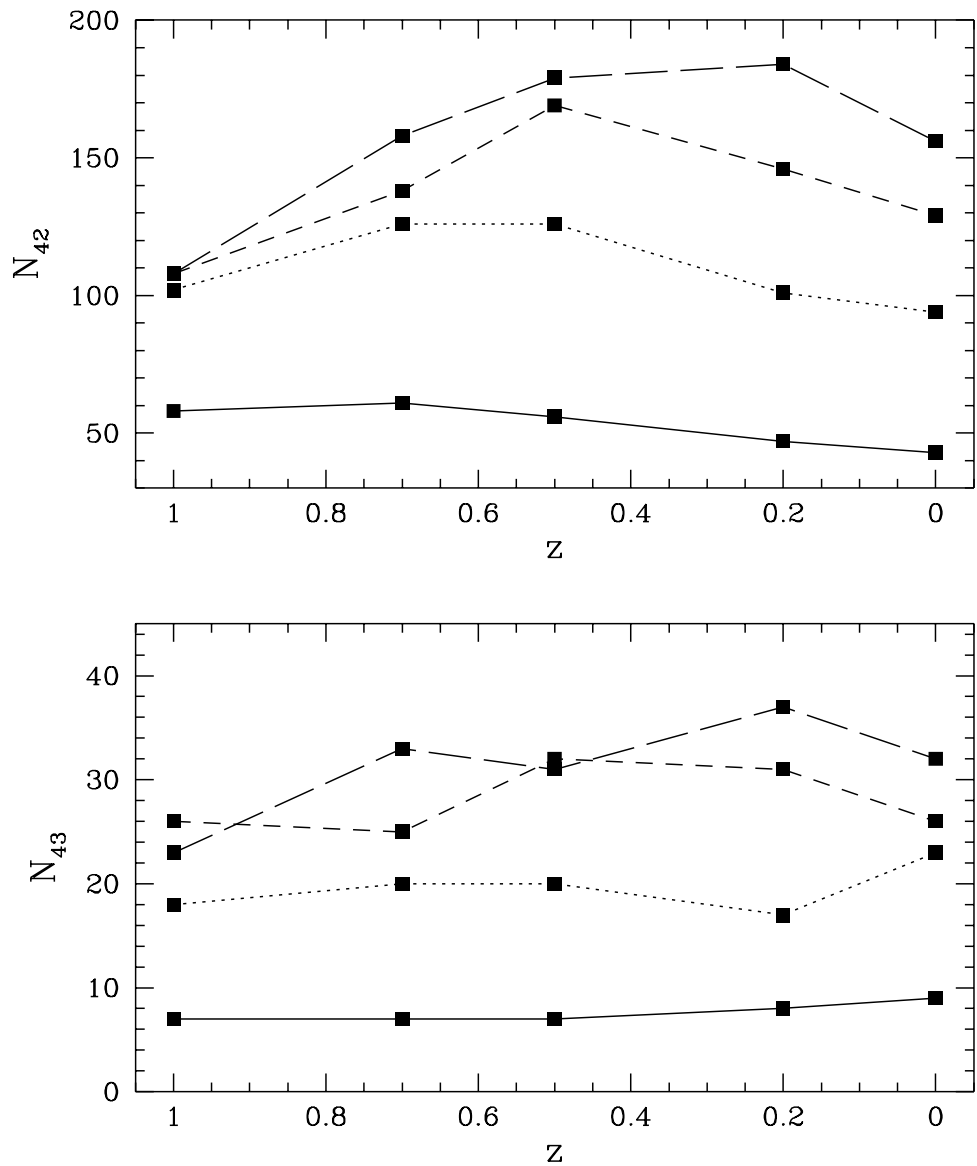,width=17.0cm,height=17.0cm}} 
\caption{
The number of clusters with luminosity $L_x\ge 10^{42}$ ($N_{42}$, upper panel)
and $L_x\ge 10^{43}$ erg\ s$^{-1}$ ($N_{43}$, bottom panel) as a function of
the redshift $z$ for the different models: BM05 (solid line), BM10 (dotted
line), BM15 (short-dashed line) and BM20 (long-dashed line). 
}
\end{figure*} 

\newpage
\begin{figure*} 
\centerline{\psfig{file=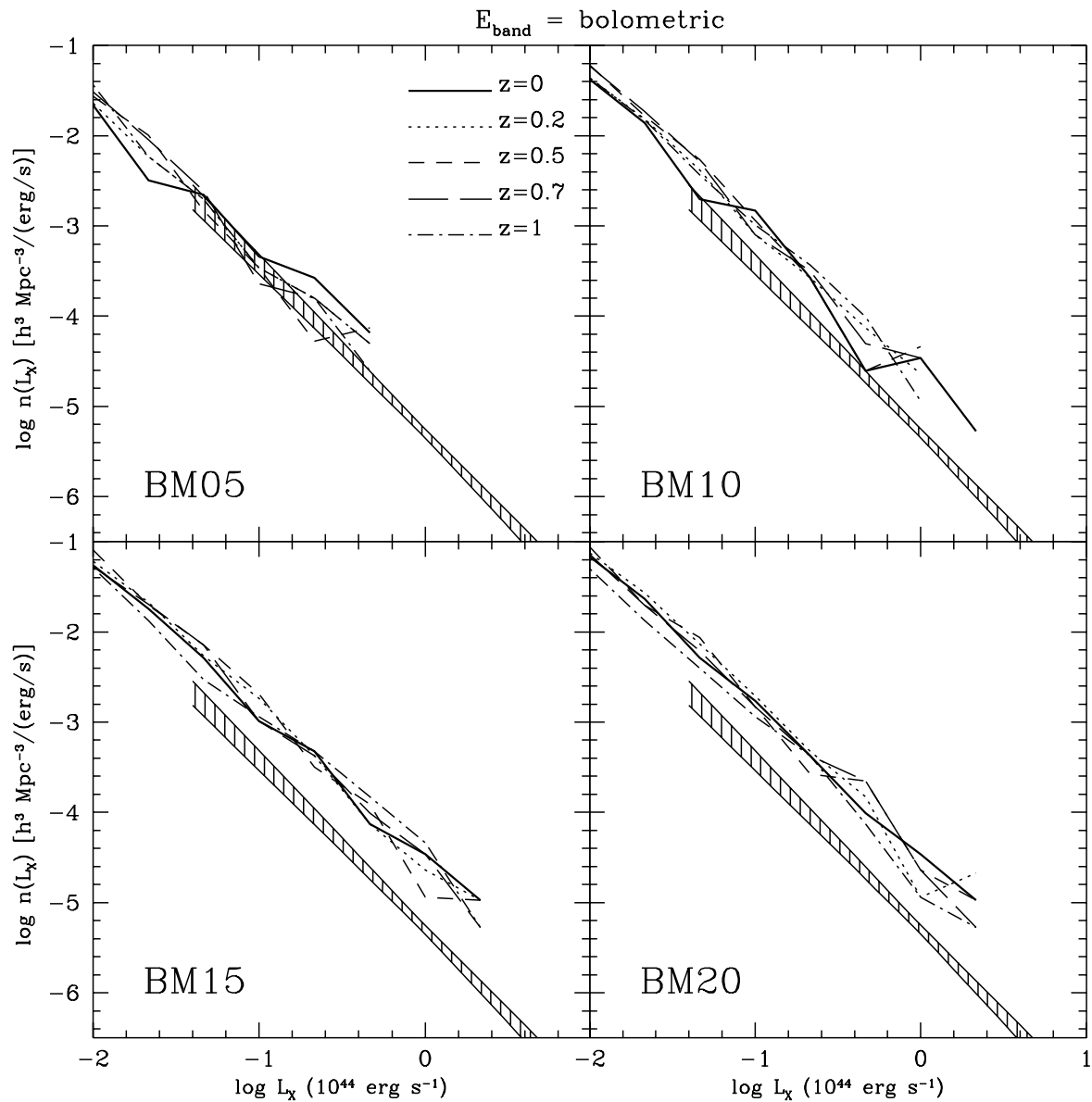,width=17.0cm,height=17.0cm}} 
\caption{
The cluster bolometric luminosity function for the different models: BM05 (top
left), BM10 (top right), BM15 (bottom left) and BM20 (bottom right). The
different curves refer to various redshift: $z=0$ (solid line), $z=0.2$ (dotted
line),  $z=0.5$ (short-dashed  line), $z=0.7$ (long-dashed line),  $z=1$
(dotted-dashed line). The hatched region shows the observational results (with
1-$\sigma$ errorbars) obtained by Ebeling et al. (1997). 
}
\end{figure*} 

\newpage
\begin{figure*} 
\centerline{\psfig{file=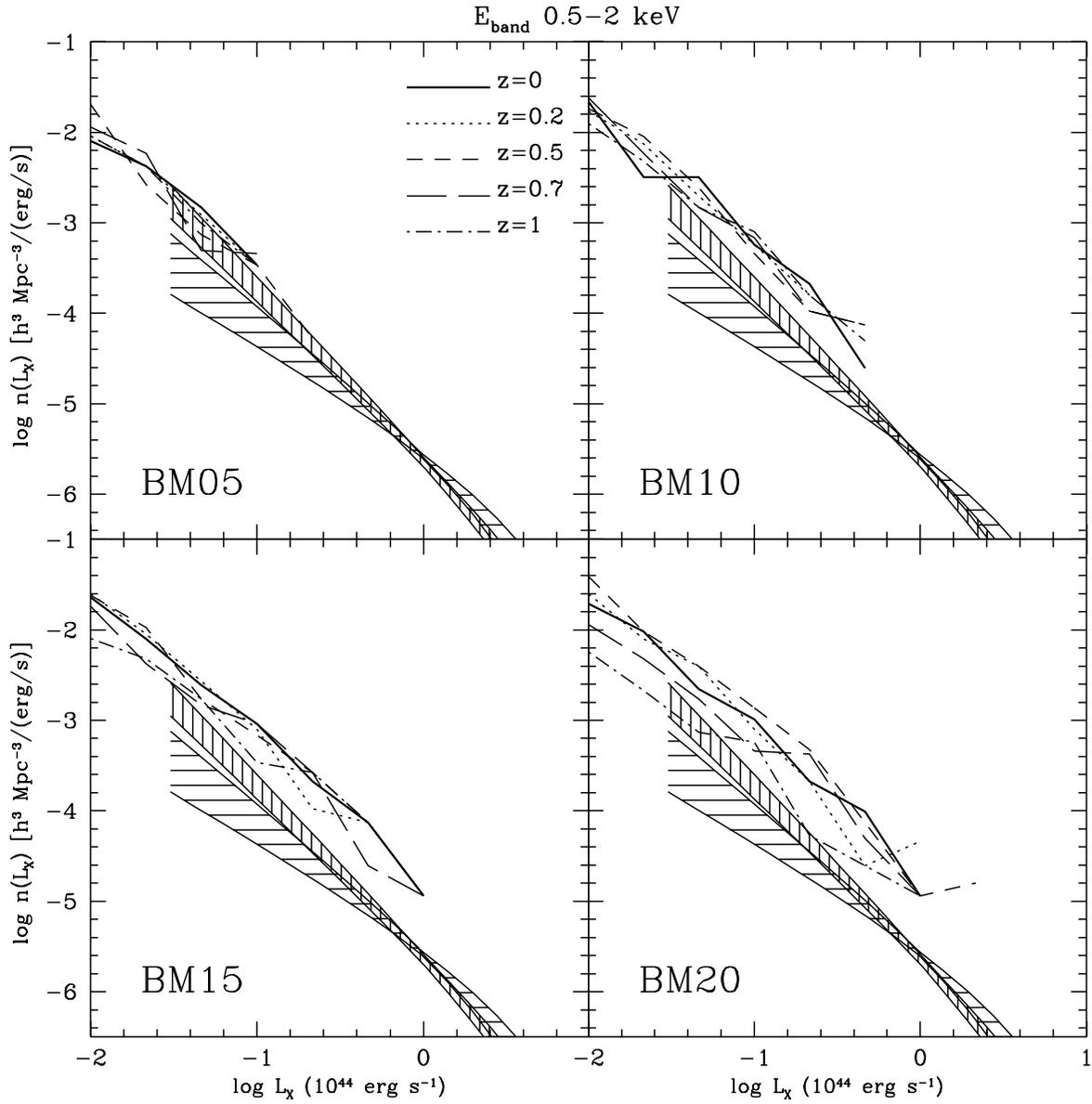,width=17.0cm,height=17.0cm}} 
\caption{
The same as Figure 5 but in the [0.5--2] keV energy band. The vertically
hatched region shows the observational results (with 1-$\sigma$ errorbars)
obtained by Ebeling et al. (1997), while the horizontally hatched one refers to
the De Grandi (1996) results. 
}
\end{figure*} 

\newpage
\begin{figure*} 
\centerline{\psfig{file=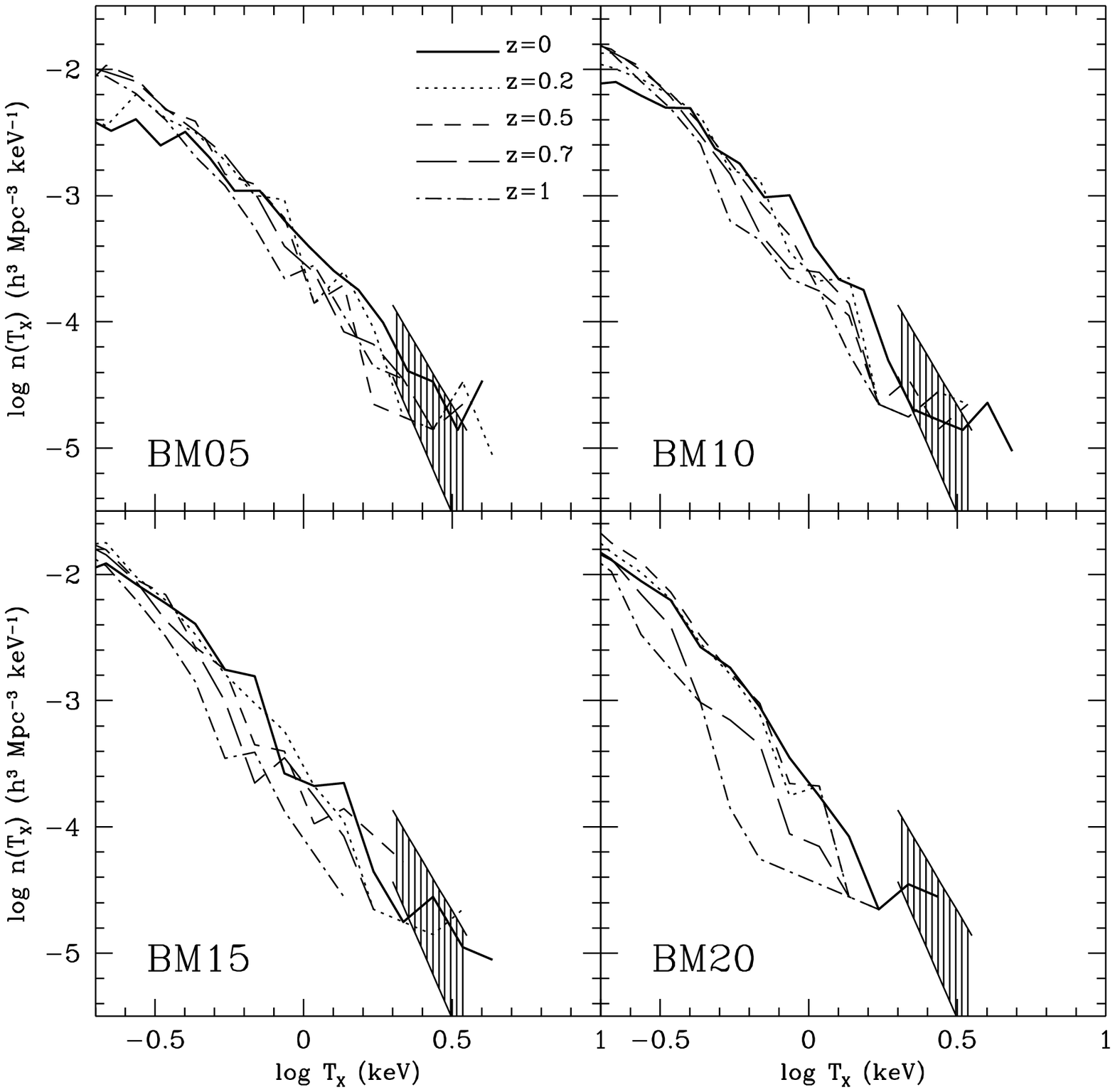,width=17.0cm,height=17.0cm}} 
\caption{
The cluster temperature function for the different models: BM05 (top left),
BM10 (top right), BM15 (bottom left) and BM20 (bottom right). The different
curves refer to various redshift: $z=0$ (solid line), $z=0.2$ (dotted line),
$z=0.5$ (short-dashed line), $z=0.7$ (long-dashed line), $z=1$ (dotted-dashed
line). The hatched region shows the observational results (with 1-$\sigma$
errorbars) obtained by Henry \& Arnaud (1991). 
}
\end{figure*} 

\newpage
\begin{figure*} 
\centerline{\psfig{file=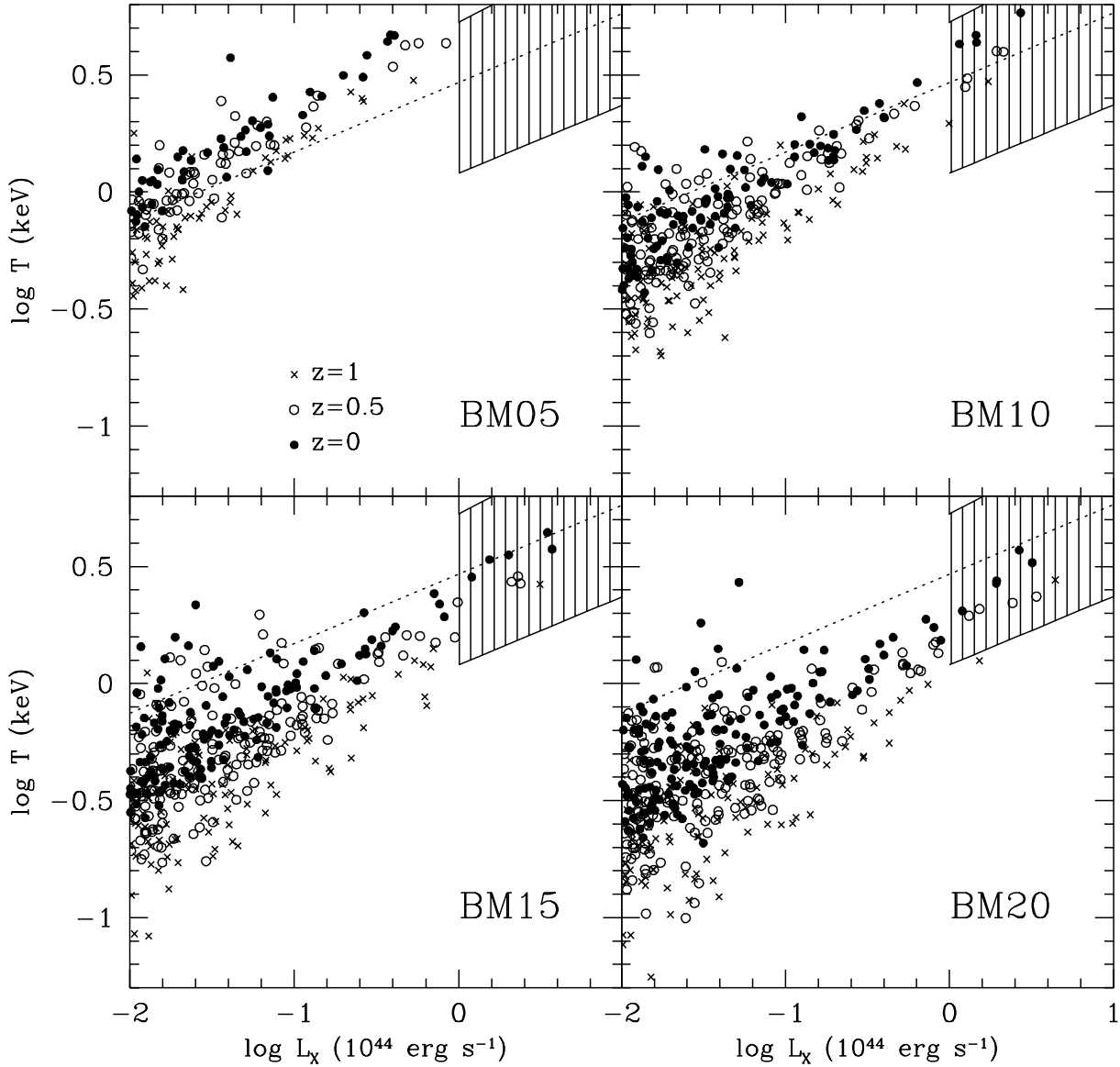,width=17.0cm,height=17.0cm}} 
\caption{
Scatter plots of the bolometric X-ray luminosity $L_x$ and the
emission-weighted temperature $T$ of the clusters  for the different models:
BM05 (top left), BM10 (top right), BM15 (bottom left) and BM20 (bottom right).
Different redshifts are displayed by different symbols: $z=1$ (crosses),
$z=0.5$ (open circles) and $z=0$ (filled circles). The hatched region shows the
observational results (with 1-$\sigma$ errorbars) obtained by Henry \& Arnaud
(1991); the dotted line refers to the fit obtained from the combined sample of
David et al. (1993). In this last case the errorbars are not explicitly
reported, but they are of the same order of magnitude of the previous ones.} 
\end{figure*} 

\end{document}